# A look at Einstein's clocks synchronization

**Nilton Penha**
Departamento de Física, Universidade Federal de Minas Gerais, Brasil.
nilton.penha@gmail.com

**Bernhard Rothenstein**
Politehnica University of Timisoara, Physics Department, Timisoara, Romania.
brothenstein@gmail.com

**Abstract**
While Einstein's clocks synchronization process is performed, one has a well defined region in which the clocks are synchronized and another one in which the clocks are not yet synchronized. The frontier between them evolves differently from the perspective of observers in relative motion. A discussion is conducted upon direct "observation" of the phenomenon and Minkowski diagrams.

**Introduction**

Special theory of relativity relies upon two postulates, stated by Einstein in his famous 1905 paper[1]. The first is the so called principle of relativity which asserts that the laws of physics hold the same in every inertial reference frame. This leads to the important outcome that no experiment in one inertial frame can distinguish it, intrinsically, from any other. The second postulate asserts that "light is always propagated in empty space with a definite velocity $c$ which is independent of the state of motion of the emitting body." Einstein's strongest justification for this postulate came from Maxwell's electrodynamics. That theory had identified light with waves propagating in an electromagnetic field and concluded that just one speed was possible for them in empty space, $c = 2.99792458 \times 10^8$ m/s, the updated value, no matter the motion of the emitter.

An event is any physical occurrence taking place at a given point in space at a given instant of time. To establish the coordinates of an event one should choose a reference frame. Let $K$ be a reference frame according to which the spatial position of an arbitrary event is given by cartesian coordinates $(x,y)$, in the case of a two-dimensional space, for example. The instant of time $ct$ at which the event happens is the reading of a clock placed exactly at its spatial position $(x,y)$. So $(x,y,ct)$ are the coordinates of the referred event according to the reference frame $K$. Similarly, if the reference frame is $K'$, the coordinates of the same arbitrary event are $(x',y',ct')$ where $ct'$ is the reading of a clock placed at spatial position $(x',y')$ in $K'$.

Notice that we treat time as $ct$ instead of $t$. This has the advantage of improving the transparency of the symmetry that exists between space and time in special relativity. We can see this clearly in the Lorentz transformations equations. It worth mention the famous assertive[2] from Herman Minkowski: "Henceforth space by itself, and time by itself, are doomed to fade away into mere shadows, and only a kind of union of the two will preserve an independent reality".

Formally, one can represent a given event as $E(x,y,ct)$ or $E'(x',y',ct')$ depending on it is described in a $K$ or $K'$ frame. The totality of possible events constitutes what is called a spacetime and $x,y,ct$ and $x',y',ct'$ are spacetime coordinates according to two different reference frames. One set of spacetime coordinates can be mapped into the other through the so called Lorentz transformations.

Suppose that you, the reader, are in the inertial frame $K$, at its spatial origin $(x,y,ct) = (0,0,ct)$ and some friend of yours is in $K'$, also at its origin $(x',y',ct') = (0,0,ct')$ which moves away with constant speed $V$, along $x$-axis. The $x'$-axis and $y'$-axis are assumed to have the same direction as the $x$-axis and $y$-axis, respectively. You may consider yourself that the inertial frame $K$ where you are in is stationary while your friend is in a non-stationary frame $K'$. Since $K$ and $K'$ are both inertial frames, your friend may also, by himself (herself), consider that he (she) is in a stationary frame $K'$ and, judge that you are in a non-stationary inertial frame $K$.

In every inertial frame it is highly convenient to imagine a standard lattice of stationary observers as small as they can be, each one with a standard clock, being all the clocks alike and stationary. You are one of those stationary observers, the one who sits at origin of *K*. It is common, among physicists, to refer figuratively to such clocks as *wristwatches*. We will follow suit here. It is also convenient to have all the wristwatches synchronized. To achieve such synchronization, one can use a procedure proposed by Einstein which goes as follows.

Consider first the *K* frame.

Initially all the wristwatches are stopped. And then one should choose a master wristwatch, usually the one at the spatial origin, the place where you are. Your wristwatch should be set to start running at an arbitrary time $ct_o$ and all the others set to start running at time $ct = ct_0 + r$ where $r$ is the spatial distance from you. Then a light source previously placed at your place emits a pulse when it starts running at time $ct_0$. The pulse propagates through the lattice and triggers off each one of the wristwatches which start reading $ct = ct_o + r$ as previously settled. Once all the wristwatches are running, they are all synchronized. But, while this process is not finished, all the lattice wristwatches, including yours, inside a circle of radius

$$r = \sqrt{x^2 + y^2} = ct \qquad (1)$$

are synchronized and those outside the circle are not synchronized yet; actually they are all stopped, by construction, waiting for the light pulse to reach them. So one has a circular frontier between the synchronized and unsynchronized wristwatches regions; such frontier propagates outwards, with constant speed *c*, from the spatial origin.

Consider now the *K'* frame.

All the wristwatches in *K'* should be initially stopped. A master wristwatch should usually be chosen as the one at the spatial origin of *K'*, which means your friend's wristwach. It should be set to start running at an arbitrary time $ct_o'$ and all the others set to start running at time $ct' = ct_0' + r'$ where $r'$ is the spatial distance from your friend's wristwatch. Then a light source previously placed at the master wristwatch lattice site emits a pulse at time $ct_0'$. The pulse propagates through the lattice and triggers off each one of the wristwatch which start reading $ct' = ct_o' + r'$ as previously settled. Once all the wristwatches are running they are all synchronized. But, while this process is not finished, all the lattice wristwatches, including that of your friend, inside a circle of radius

$$r' = \sqrt{x'^2 + y'^2} = ct', \qquad (2)$$

are synchronized and those outside the circle are not synchronized yet; actually they are all stopped, by construction, waiting for the light pulse to reach them. So one has a circular frontier between the synchronized and unsynchronized wristwatches regions; such frontier propagates outwards, with constant speed *c*, from the spatial origin.

Important: with no loss of generality one can assume that $ct_0 = ct_0' = 0$ and that both origins *O* and *O'* are coincident, i.e., $(x,y,ct) = (0,0,0) = (x',y',ct')$ when a *unique* light source emits a pulse which propagates through *K* and *K'* synchronizing all the wristwatches. According to the second postulate it does not matter whether the light source is at rest in *K* or at rest in *K'*. Such event may be represented by $E_O(0,0,0)$ and $E'_{O'}(0,0,0)$ according to K and K' frames.

**Light clocks**

A light clock is a device made of two parallel mirrors, say, $M_O$ and $M$, just in front of each other separated by a fixed distance *d*. A light source placed at one of the mirrors emits a pulse which should be bouncing between them. It is common to refer to a clock by mentioning its *tick-tacks*. Let the *tick* be moment at which the light pulse is at the position of $M_O$ mirror and let the *tack* be the moment at which the light pulse is at the position of the mirror *M*. The elapsed time between two consecutive *ticks* in the same given light clock should be a characteristic of such light clock. Such period is what one usually calls a proper time. Let us represent such proper time as $2c\tau_d = 2d$. Since the speed of light does not depend on the direction, $c\tau_d$ is the elapsed proper time for both up and down propagation of the light ray inside the light clock.

Consider a light clock at rest in the inertial reference frame *K*. Let us call it *K light clock*. Let $M_O$ be placed at the origin *O* such that its plane is perpendicular to *y*-axis and *M* placed at $(x,y,ct) = (0,d,ct)$ parallel to the first. You, along with your wristwatch, are at the position of the $M_O$ mirror.

Consider also another of the same light clock, at rest in the inertial reference frame *K'*. Let us call it *K' light clock*. Let the mirror $M'_{O'}$ be placed at the origin *O'* such that its plane is perpendicular to *y'*-axis and *M'* placed at $(x',y',ct') = (0,d',ct')$ parallel to the first. Your friend, along with his (her) wristwatch, are at the position of the $M'_{O'}$ mirror. The proper period in this case is $2c\tau'_{d'} = 2d'$. Assuming that both light clocks are alike, *d'*, the mirrors spatial separation while the *K'* clock is at rest, should be taken equal to *d,* the same separation, the mirrors separation while the *K* clock is at rest*,* and consequently, they both have a same proper time ($c\tau'_{d'} = c\tau_d$).

In particular we should be interested in comparing the elapsed time between two given events as 'seen' by you at rest in *K* and by your friend at rest in *K'*. Again notice that the clocks are made alike. They are made to measure the time in the same way. In fact, if they were side by side at rest with respect to each other one would not detect difference in their measurements.

## The viewpoints of *K* and *K'*

Let us return to above synchronization procedure and let *O* and *O'* be the master wristwatches of *K* and *K'* located at their respective spatial origins *O* and *O'*. They are respectively yours and your friend's wristwatches.

Let us consider the moment at which the synchronization procedure starts for both inertial frames and consider your point of view. For our purposes here let the mirrors *M* and *M'* to be half-silvered.

See Figure 1. Light is emitted at the common spatial origins *O* and *O'*, where there are the mirrors $M_O$ and $M'_{O'}$ when wristwatches at such spatial position read $ct_0 = 0 = ct'_0$ by definition. This corresponds to a *K* and *K'* light clocks *tick*.

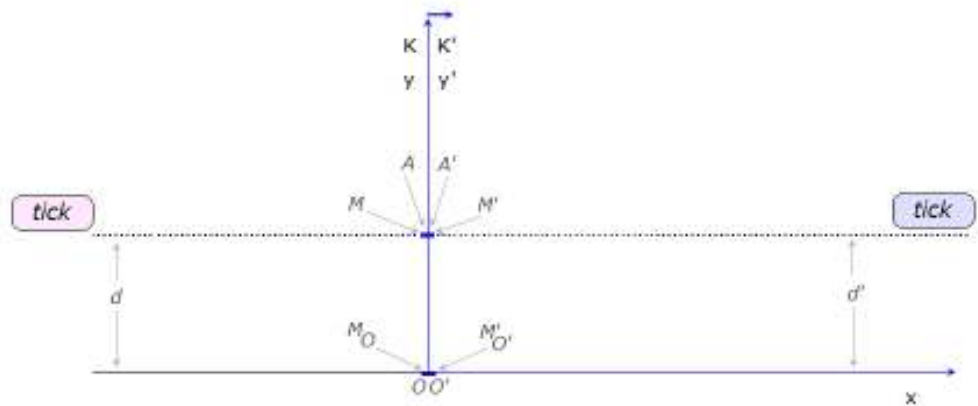

**Figure 1** – Your point of view in *K*. While *K* is assumed stationary, *K'* is moving right with constant speed $V = \beta c$ at time $ct_0 = 0 = ct'_0$ when a light pulse is emitted at the common origins *O* and *O'*. All wrist watches in both frames are stopped.

See Figure 2. Light reaches position *A* where there are a half-silvered mirror *M* and a wristwatch *A,* which just reads $ct_A = c\tau_d$. Let such event be represented by $E_A(x_A,y_A,ct_A) = E_A(0,d\ c\tau_d)$. This corresponds to a *K* light clock *tack*. All wristwatches at rest in *K* which are inside a circle of radius $r_A = c\tau_d$ show the same time reading $c\tau_d$ and those outside the circle are still stopped (idling).

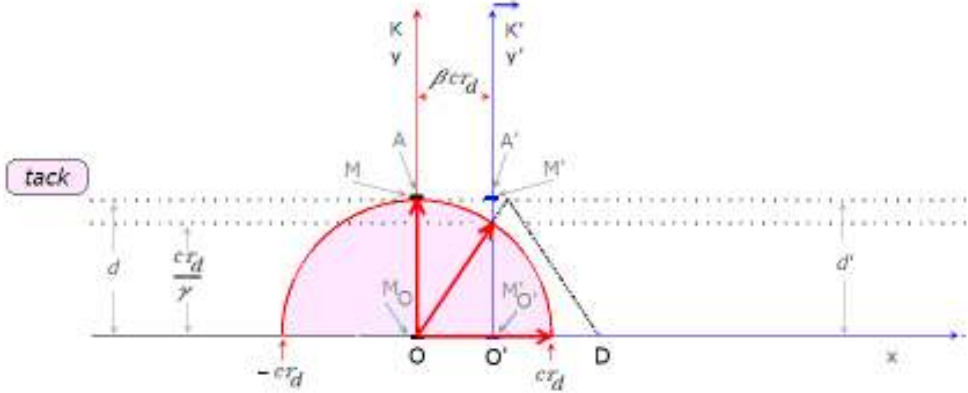

**Figure 2** – Your point of view in *K*, at the moment the light wavefront reaches the half-silvered mirror *M* and wrist watch *A*, both at rest in *K*. All wristwatches at rest in *K* which are inside a circle of radius $r_A = c\tau_d$ show the same time reading $c\tau_d$. All wristwatches outside the circle in *K* are still stopped.

### Time dilation

See Figure 3. Light emitted at *O* (and *O'*) keeps expanding its wavefront and reaches the half-silvered mirror *M'* at which position there is a wristwatch *A'*, both at rest in *K'*, and the wristwatch *B*, at rest in *K*. Such event may be represented by $E'_{A'}(x'_{A'}, y'_{A'}, ct'_{A'}) = E'_{A'}(0, d', c\tau'_{d'})$, in *K'*, and by $E_B(x_B, y_B, ct_B) = E_B(\beta ct_B, d, ct_B)$, in *K*. The wristwatches *A'* and *B*, they momentarily face each other; while *B* reads $ct_B > c\tau_d$, *A'* just reads $ct'_{A'} = c\tau'_{d'} = c\tau_d$. This corresponds to a *K'* light clock *tack*. Also, light previously reflected at the mirror *M* is going on its way back the origin *O*, the place where you are still. All wristwatches at rest in *K* which are inside a circle of radius $r_B = ct_B$ show the same time reading $ct_B$ and those outside the circle are still stopped.

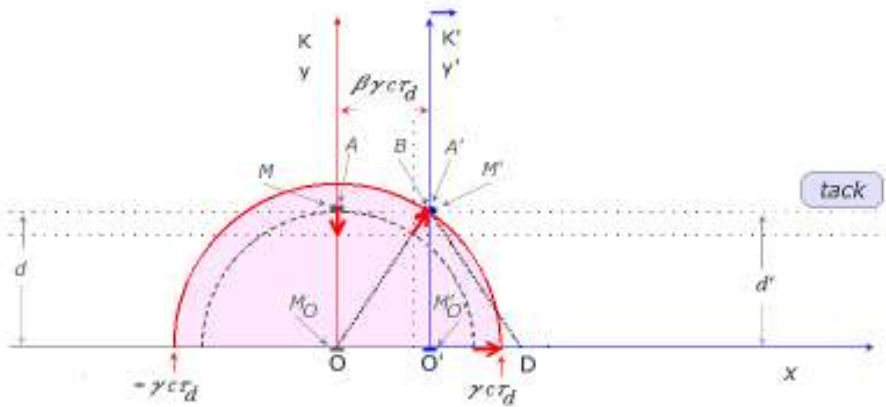

**Figure 3** – Your point of view in *K*, at the moment the light wavefront reaches the half-silvered mirror *M'* and wristwatch *A'*, both at rest in *K'*, and wristwatch *B*, which is at rest in *K*; *B* and *A'* are just facing each other at the moment the light pulse strikes them. All wristwatches at rest in *K* which are inside a circle of radius $r_B = ct_B$ show the same time reading $ct_B$. All wrist watches outside the circle in *K* are still stopped.

A simple calculation, based on Pithagoras Theorem, leads to

$$ct_B = \gamma\, c\tau_d \qquad (3)$$

where

$$\gamma = \frac{1}{\sqrt{1-\beta^2}}. \qquad (4)$$

Since $c\tau'_{d'} = c\tau_d$ (by construction), we have

$$ct_B = \gamma\, c\tau'_{d'} \qquad (5)$$

This means that the elapsed time between same two events which is measured by your friend in $K'$ seems larger to you in K. To you the elapsed time is larger by the $\gamma$ factor. Although the light clocks are made alike they show different readings when they are moving with respect to the observer. This is what one calls *time dilation*:

$$ct = \gamma\, ct' \qquad (6)$$

**Length contraction**

At this point we assume the existence of a rigid *rod*, extended along the *x'-axis,* at rest in $K'$, such that the left end is at $O'(0,0,ct')$ and the right one is at $C'(x'_{C'},0,ct')$. Let its length be $(x'_{C'} - 0) = d'$ as measured by your friend. He (she) actually can measure it by using a standard meter. So the still rod has a proper length equal to

$$L' = c\tau'_{d'} \qquad (7)$$

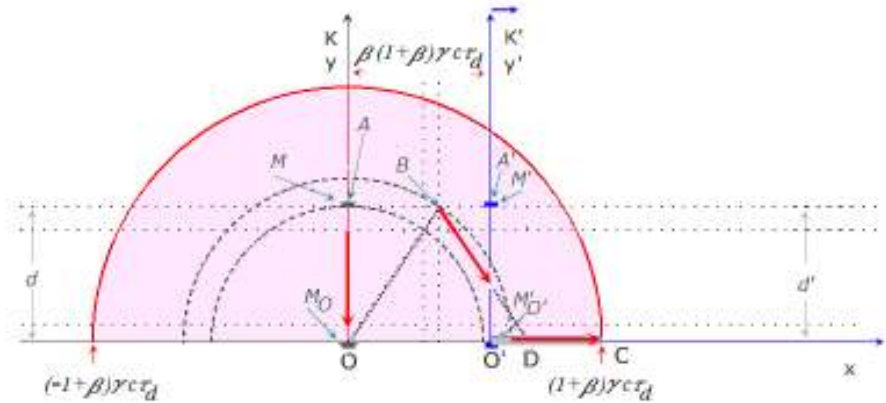

**Figure 4** – Your point of view, in $K$, at the moment the light propagating directly from $O$ ($O'$) reaches the right end of the rod. According to you light spends a time equal to $(1+\beta)\gamma c\tau_d$ to perform the way to the right end of the rod. All wristwatches at rest in $K$ which are inside a circle of radius $r_C = (1+\beta)\gamma c\tau_d$ show the same time reading $(1+\beta)\gamma c\tau_d$. All wristwatches outside the circle in $K$ are still stopped.

See Figure 4. Part of light reflected at mirror $M'$ follows its way to the mirror $M'_{O'}$ at the spatial origin $O'$ of $K'$ while this one moves along *x-axis* with dimensionless speed $\beta$. Also the light ray propagating along *x-axis,* from $O(0,0,0)$ (and $O'(0,0,0)$) reaches the right end of rod (that we assumed to exist at rest in $K'$) which has a proper length $L' = c\tau'_{d'}$. At the moment light reaches the rod right end, a wristwatch $C$, at rest in $K$, show the reading $ct_C$,

$$ct_C = ct_B + \beta\, ct_B = \gamma\, c\tau_d + \beta\, \gamma\, c\tau_d \qquad (8)$$

while $C'$, at rest in $K'$, which is momentarily just face to face with $C$, shows $ct'_{C'}=(1+\beta)c\tau'_{d'}$. To you, the observer at rest in $K$, light runs through a distance, along x-axis, just equal to

$$x_C - x_O = (1+\beta)\gamma c\tau_d \qquad (9)$$

before reaching the right end of the rod at $(x_C,y_C,ct)=((1+\beta)\gamma c\tau_d, 0, (1+\beta)\gamma c\tau_d)$. Let us refer to such event as $E_C(x_C,y_C,ct_C) = E_C((1+\beta)\gamma c\tau_d, 0, (1+\beta)\gamma c\tau_d)$ when described in $K$ and as $E'_{C'}(x'_{C'},y'_{C'},ct'_{C'}) = E'_{C'}((1+\beta)c\tau'_{d'}, 0, (1+\beta)c\tau'_{d'})$ when described in $K'$. While light propagates from $O$ directly to $C$, the left end of the rod moves to $(\beta(1+\beta)\gamma c\tau_d, 0, (1+\beta)\gamma c\tau_d)$ along the x-axis. The spatial distance between these two points is the length that the rod appears to have to you:

$$L = ct_C - \beta ct_C = (1+\beta)\gamma c\tau_d - \beta(1+\beta)\gamma c\tau_d \qquad (10)$$
$$L = (1-\beta^2)\gamma c\tau_d = \sqrt{(1-\beta^2)}\, c\tau'_{d'} \qquad (11)$$
$$L = \frac{1}{\gamma}L'. \qquad (12)$$

The rod looks shorter if it belongs to moving inertial frames. This is the so called relativistic *length contraction*.

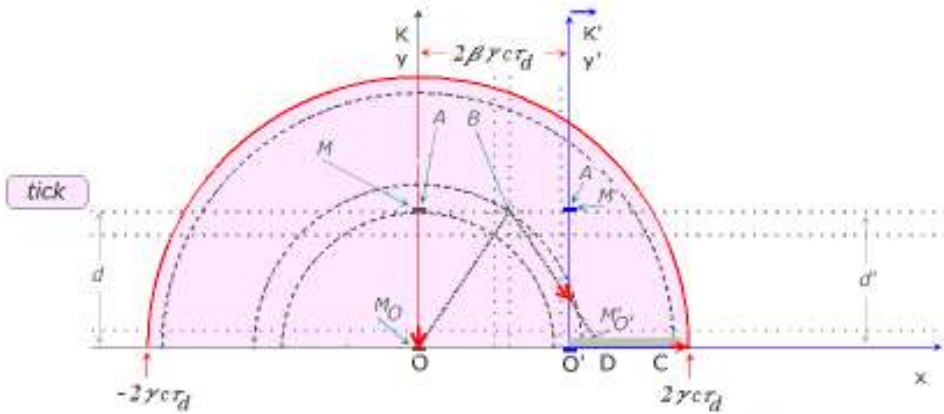

**Figure 5**– Your point of view, in $K$, at the moment the light reflected by mirror $M$ reaches back your position. This corresponds to a *tack* in your reference frame $K$ and closes a cycle. All wristwatches outside the circle in $K$ are still stopped.

To understand the viewpoint of your friend you just have to use $-\beta$ instead of $\beta$, and change primed for unprimed variable. You will see that he (she) also thinks that your clock works slower (you are moving and he (she) is at rest); also if you have a rod at rest in $K$ it will seem shorter to him (her). The situation is symmetric.

**Lorentz Transformations**

From what is discussed above it is possible to infer the so-called Lorentz Transformations. From (12) one can write

$$(x_C - \beta ct_C) = \frac{1}{\gamma}(x'_{C'} - 0) \tag{13}$$

Although the proper length of above rigid rod was assumed to be $c\tau'_{d'}$ it actually could have any length and the conclusion about the length contraction would have been the same. So the position *C'* mentioned above is arbitrary. Then expression (13) can be put in the following way:

$$x' = \gamma(x - \beta ct). \tag{14}$$

This gives the spacetime coordinate *x'* in terms of spacetime coordinates *x* and *ct*. If one change $\beta$ for $-\beta$ and primed for unprimed coordinates one gets

$$x = \gamma(x' + \beta ct'). \tag{15}$$

Now if we insert (14) into (15) one has

$$x = \gamma(\gamma(x - \beta ct) + \beta ct'), \tag{16}$$

$$\beta\gamma ct' = (1 - \gamma^2)x + \beta\gamma^2 ct, \tag{17}$$

$$ct' = \gamma(ct - \beta x). \tag{18}$$

This expression gives spacetime coordinate *ct'* in terms of spacetime coordinates *x* and *ct*. Again by changing $\beta$ for $-\beta$ and primed for unprimed one has

$$ct = \gamma(ct' + \beta x'). \tag{19}$$

Expressions (14) and (18) together with $y = y'$ are the Lorentz Transformations

$$x' = \gamma(x - \beta ct) \tag{20}$$
$$y' = y \tag{21}$$
$$ct' = \gamma(ct - \beta x), \tag{22}$$

and expressions (15) and (19) are the inverse Lorentz Transformations

$$x = \gamma(x' + \beta ct') \tag{23}$$
$$y = y' \tag{24}$$
$$ct = \gamma(ct' + \beta x'). \tag{25}$$

**Important:** *By exchanging x' for ct' and x for ct in the Lorentz Transformation (inverse Lorentz Transformation) it becomes clear the symmetry between space and time spacetime coordinates* (refer to Minkowski).

### Again the viewpoint of K and K'

In the process of synchronization, while the pulse propagates radially triggering off all the wristwatches on its way one has an inner region where the wristwatches are all synchronized and an outer region where the wristwatches are not yet synchronized. To you, at rest in the inertial frame *K*, the frontier between the mentioned regions propagates with no distortion at constant speed *c* according to second postulate of special relativity and has the shape of a circle. The radius *r* of such circular frontier is given by expression (1). To your friend at rest in *K'*, the circular frontier has a radius *r'* that satisfies expression (2). However the frontier propagation that happens in *K'* appears to you distorted; your friend also "sees" the frontier propagation that happens in your frame as distorted.

Let the set of simultaneous (*fixed ct'*) events happening on the circular frontier according to observers at rest in *K'* be represented by $E'(x', y', ct')$ in Cartesian coordinates, and $E'(r'\cos\theta', r'\sin\theta', ct')$ in polar coordinates as well. By applying appropriately Lorentz transformations to the simultaneous events spacetime coordinates in *K'*, and setting $r' = ct'$ one should get how these same events are "seen" by stationary observers in *K*:

$$E(x, y, ct) = E(r\cos\theta, r\sin\theta, ct) \tag{25}$$

where

$$x = r\cos\theta = \gamma r'(\cos\theta' + \beta), \tag{26}$$
$$y = r\sin\theta = r'\sin\theta', \tag{27}$$
$$ct = \gamma r'(1 + \beta\cos\theta'). \tag{28}$$

The polar coordinates *r*, *θ* in *K* are explicitly expressed as

$$r = \gamma r'(1 + \beta\cos\theta'), \tag{29}$$
$$\theta = \arctan\left(\frac{\sin\theta'}{\gamma(\cos\theta' + \beta)}\right). \tag{30}$$

The circular wavefronts in *K'* are seen as ellipses in *K*.

In the sequence of figures 6-10 we plot the *K* and *K'* frontiers between the synchronized and the not yet synchronized regions as "seen" by you, at rest observer in *K*, for the same different stages shown by figures 1-5. All the wristwatches inside the circles are already synchronized in *K* frame. All the wristwatches inside the ellipses are already synchronized in *K'*. In the overlapped regions all the wristwatches in *K* show the same reading *ct* and those in *K'* show the reading *ct'*; the relation between the readings is $ct = \gamma ct'$. On the left upper corner of each figure we schematically show the wristwatches readings for both frames.

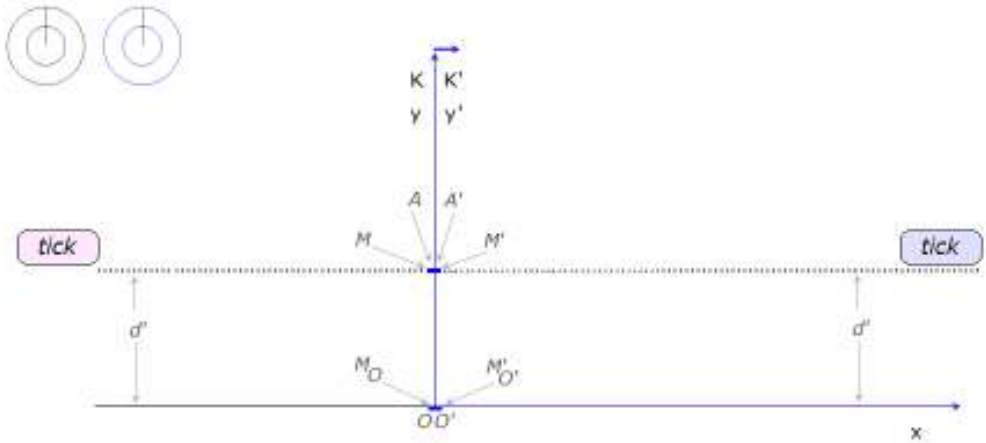

**Figure 6** – Just for an easy comparison with the figures ahead we show here the scenario at $ct = 0 = ct'$.

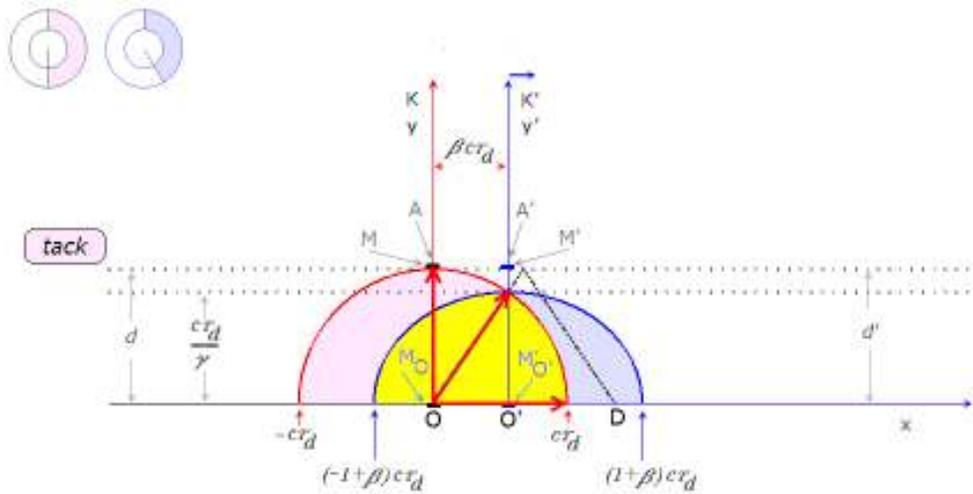

**Figure 7** – Your point of view in $K$, at the moment the light wavefront reaches the half-silvered mirror $M$ and wristwatch $A$, both at rest in $K$. All wristwatches at rest in $K$ which are inside a circle of radius $r_A = c\tau_d$ show the same time reading $c\tau_d$ and those outside the circle in $K$ are still stopped. The circular wavefront in $K'$ is "seen" as an ellipse in $K$. Those wristwatches in $K'$ which are already synchronized should be showing $c\tau_d/\gamma$.

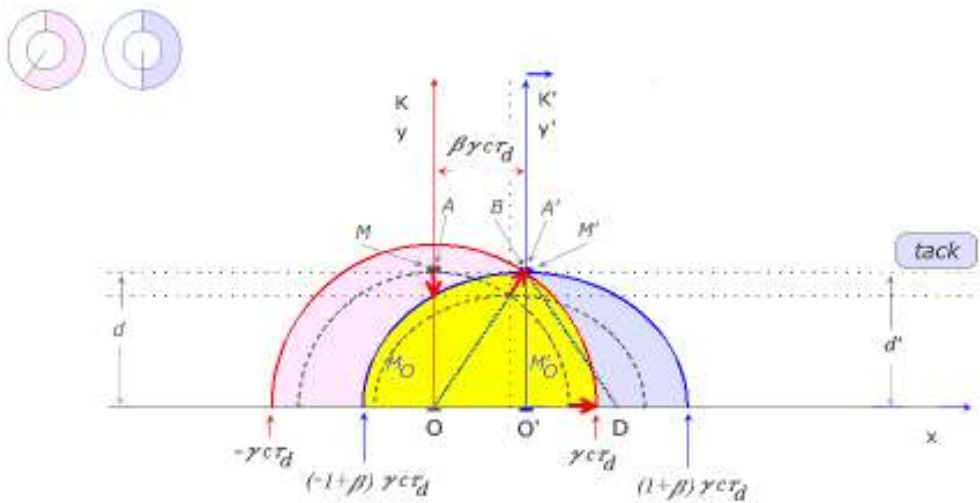

**Figure 8** – Your point of view in $K$, at the moment the light wavefront reaches the half-silvered mirror $M'$ at rest in $K'$ and wristwatch $A$, at rest in $K$. All wrist watches at rest in $K$ which are inside a circle of radius $r_A = \gamma c\tau_d$ show the same time reading $\gamma c\tau_d$ and those outside the circle in $K$ are still stopped. The circular wavefront in $K'$ is "seen" as an ellipse in $K$. Those wristwatches in $K'$ which are already synchronized should be showing $c\tau_d$.

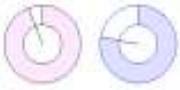
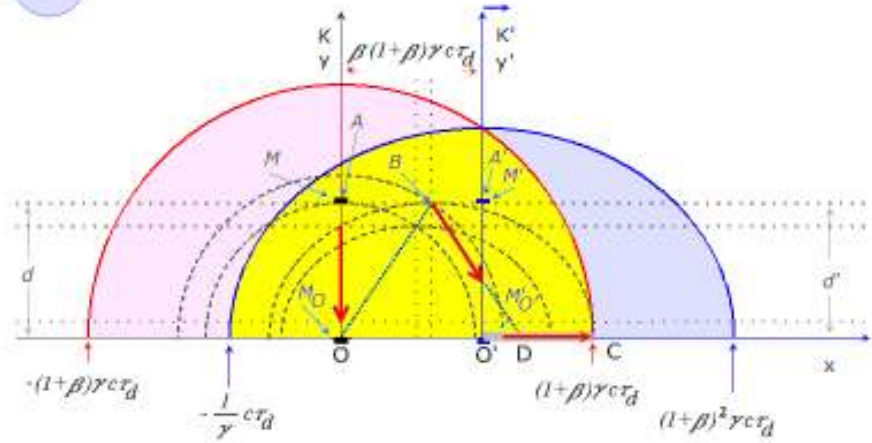

**Figure 9** – Your point of view, in *K*, at the moment the light propagating directly from *O* (*O'*) reaches the right end of the rod. According to you light spends a time equal to $(1+\beta)\gamma\tau_d$ to perform the way to the right end of the rod. All wristwatches at rest in *K* which are inside a circle of radius $r_C = (1+\beta)\gamma c\tau_d$ show the same time reading $(1+\beta)\gamma c\tau_d$. While this happens, the origin *O'* is displaced by the distance $\beta(1+\beta)\gamma c\tau_d$.

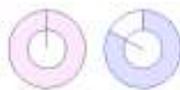
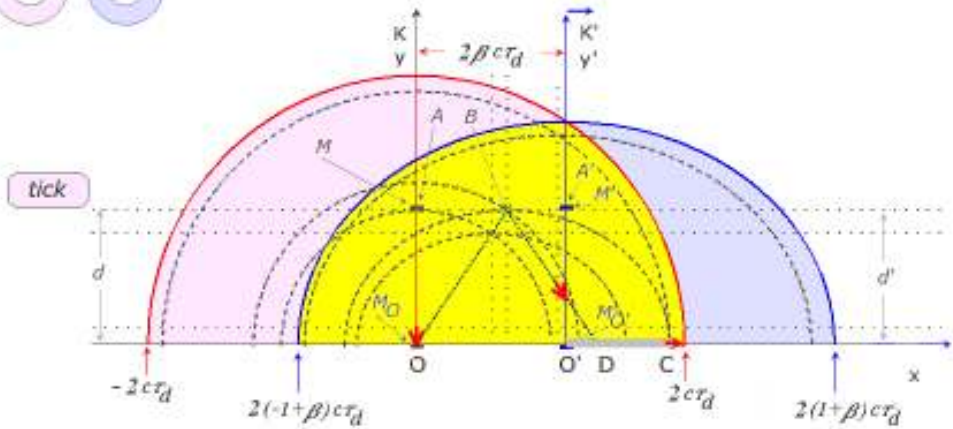

**Figure 10** – Your point of view, in *K*, at the moment the light propagating along the *y*-axis reaches back the origin *O* closing a cycle (*tick-tack-tick*). The circular wavefront in *K'* is "seen" as an ellipse in *K*.

### Minkowski diagrams

In the sequence of figures 11-15 we show the Minkowski diagrams for every stage shown in figures 5-10. There is a region in which the two lightcones overlap. All the wristwatches in *K* inside that region exhibit the reading $ct = \gamma ct'$, where $ct'$ is the reading of those already synchronized wristwatches in *K'*. The first *ticks* of the *K* and *K'* light clocks are simultaneous events, just by construction. The subsequent *ticks* and *tacks* are not simultaneous. Simultaneity is a relative concept.

**Figure 11** – Your point of view, in *K*, at the moment that the synchronization procedure starts in both reference frames. ($y = 0 = y'$)

**Figure 12** – Your point of view, in *K*, at the moment the light propagating directly from *O (O')* reaches mirror *M*, which is at rest in *K*. All wristwatches at rest in *K* which are inside a circle of radius $r = c\tau_d$ show the same time reading $c\tau_d$. Those wristwatches in *K'* which are synchronized show $c\tau_d/\gamma$. The *K* and *K'* wrist watches can only be compared in the overlapped region which goes from $x = (-1+\beta)\, c\tau_d$ to $x = c\tau_d$. ($y = 0 = y'$)

**Figure 13** – Your point of view, in *K*, at the moment the light propagating directly from *O (O')* reaches *M'* at rest in *K'*. One can see that to your friend, light has already reached the right end of the rod. To you light has not reached that point yet. This is an event which is not simultaneous to both *K* and *K'*. All wrist watches at rest in *K* which are inside a circle of radius $r = \gamma c\tau_d$ show the same reading $\gamma c\tau_d$. Those wristwatches in *K'* which are synchronized show $c\tau_d$. The *K* and *K'* wristwatches can only be compared in the overlapped region (from $x=(-1+\beta)\gamma c\tau_d$ to $x=\gamma c\tau_d$). ($y = 0 = y'$)

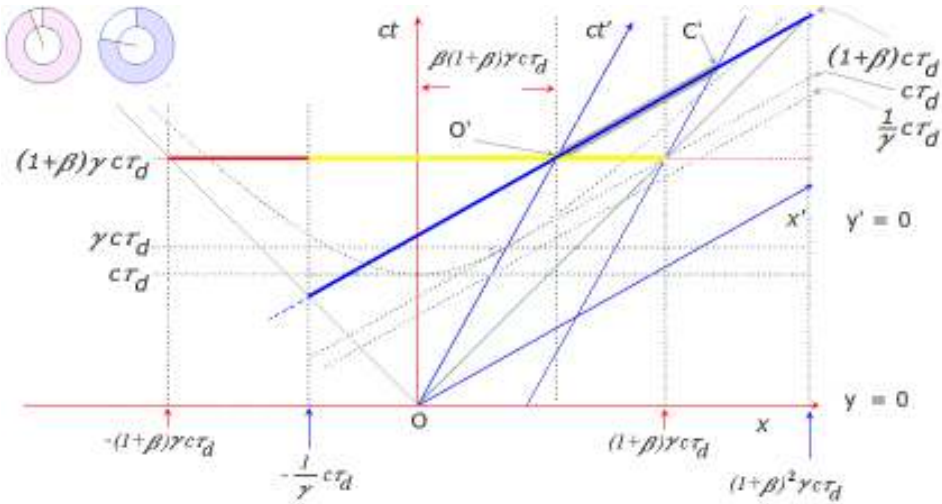

**Figure 14** – Your point of view, in $K$, at the moment the light propagating directly from $O$ ($O'$) reaches the right end of the rod ($C'$). According to you light spends a time equal to $(1+\beta)\gamma c\tau_d$ to perform the way to the right end of the rod. All wristwatches at rest in $K$ which are inside a circle of radius $r_C = (1+\beta)\gamma c\tau_d$ show the same time reading $(1+\beta)\gamma c\tau_d$. Those wristwatches in $K'$ which are synchronized show $(1+\beta)c\tau_d$. The $K$ and $K'$ wrist watches can only be compared in the overlapped region (from $x = -c\tau_d/\gamma$ to $x=(1+\beta)\gamma c\tau_d$).

The right end of the rod is reached at time $(1+\beta)\gamma c\tau_d$ according to you; all the wristwatches in $K$ which are inside a circle of radius $(1+\beta)\gamma c\tau_d$ show the same time reading $(1+\beta)\gamma c\tau_d$. Those wristwatches in $K'$ which are synchronized show $(1+\beta)c\tau_d$.

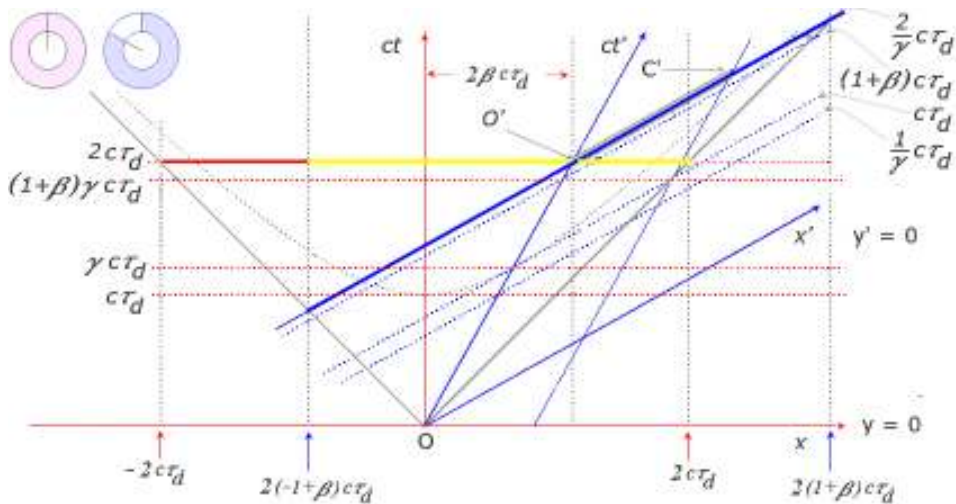

**Figure 15** – Your point of view, in $K$, at the moment the light which was reflected by mirror $M$ reaches back the origin $O$, closing a cycle (*tick-tack-tick*). The $K$ and $K'$ wrist watches can only be compared in the overlapped region (from $x = 2(-1+\beta)c\tau_d$ to $x=2c\tau_d$).

**Discussion**

We have discussed Einstein's clock synchronization as applied to two inertial reference frames with relative movement. Observer at rest in each of the frames realize they are in a stationary system and see the light propagating from the master wristwatch as wavefronts which are circular (2+1 spacetime) and so the frontier between the already synchronized wristwatches region and that of the not yet synchronized watches is also circular. When he (she) attempt to follow the frontier in the non stationary frame he (she) realizes that the referred frontier is elliptical. Pictures are plotted to explicit the evolution of such frontiers and the concept of relative simultaneity. Also with the help of two identical light clocks, one in each frame, we infer *time dilation*, *length contraction* and *Lorentz Transformations*.